\documentstyle[aps,epsfig,preprint,prl]{revtex}

\title{Reversible stretching of homopolymers and random heteropolymers}
\author{Phillip L. Geissler and Eugene I. Shakhnovich\\
{\it Department of Chemistry and Chemical Biology,
Harvard University, Cambridge, MA 02138}}

\begin{document}

\maketitle 

\begin{abstract}
We have analyzed the equilibrium
response of chain molecules to stretching.
For a homogeneous sequence of monomers,
the induced transition from compact globule to extended coil
below the $\theta$-temperature is predicted to be sharp.
For random sequences, however,
the transition
may be smoothed by a prevalence of necklace-like
structures, in which globular regions and coil regions 
coexist in a single chain.
As we show in the context of a random copolymer,
preferential solvation of one monomer type
lends stability to such structures.
The range of stretching forces over which necklaces
are stable is sensitive to chain length as well
as sequence statistics.
\end{abstract}
\newpage

Experiments probing the mechanical response of individual protein
molecules have demonstrated that certain domains can withstand
significant stretching forces before unfolding
\cite{Fernandez_nat_1998,Gaub_1998,Fernandez_nat_1999,Fernandez_PNAS_2000,Bustamante_PNAS_2000}.  
At strain rates 
much greater than unperturbed
rates of unfolding,
measured restoring forces imply that the native structures
remain largely intact up to a threshold force, 
at which they unfold to an extensible state.
Threshold forces for these domains are typically in the range
$\sim$ 10-100 $T/a$, where $T$ is temperature (in units such that
Boltzmann's constant is unity) and $a$ is an average
distance between neighboring, connected monomers in the chain.
To be sure, these experimental systems are out of equilibrium,
as highlighted by wide hysteresis upon releasing strain
and broad distributions of threshold forces.
Nevertheless, it is reasonable to expect that 
such proteins undergo sharp unfolding transitions
under adiabatic stretching conditions as well.

The microscopic origin of this mechanical strength is
not well understood.  As Wolynes and coworkers have
pointed out, small forces applied to the end-to-end distance
of a protein may couple very weakly to the reaction
coordinate for folding\cite{Wolynes_PNAS_1999}.
Atomically detailed
simulations of rapid stretching
instead suggest an important role for certain backbone
topologies that are stabilized by groups of hydrogen bonds
\cite{Schulten_proteins_1999}.
In this Letter we explore the equilibrium
force-induced transition at 
a coarse-grained level, using simple estimates of the relevant 
free energetics.  Specifically, we distinguish between aspects
of the transition
that are homopolymeric in nature, and those that arise from
the quenched disorder characterizing random heteropolymers.
We show that necklace-like structures, as depicted in 
Fig.~\ref{geometry}~(c), 
occur with low probability
in long homopolymers, but may be stabilized
over a finite range of force and temperature 
by sequence heterogeneity.
We describe the features of sequence statistics that 
affect this stability, and thereby determine
mechanical strength.
Our results may thus shed light on evolutionary
design principles for proteins whose functions are
mechanical in nature.

We consider first a chain of 
$N$ identical monomers in solution.
In the absence of stretching force and 
below the $\theta$-temperature ($\theta$), 
the chain adopts a compact, globular 
conformation (Fig~\ref{geometry}~(a)).
Because this state exhibits only small fluctuations
in monomer density, its free energy is well approximated
using mean field theory\cite{Grosberg_Khokhlov}:
\begin{equation}
F_{\rm g} \simeq B N + \gamma N^{2/3} -TS_0.
\label{globule_free_energy}
\end{equation}
Here, $B$ is the excess free energy per particle  
of a fluid of unconnected monomers at the same temperature and density
as the globule.  The second term in Eq.~\ref{globule_free_energy}
includes surface energetics of the globule-solvent
interface, 
as well as the conformational entropy $S_{\rm sph}$ lost
upon confining the chain to a spherical volume with radius 
$R\sim N^{1/3}$.  $S_0$ is the entropy of an ideal, unconfined chain.

We imagine that the principal effect of a small stretching
force, $f$, 
on the ends of a homopolymeric globule
is to deform its spherical geometry
(Fig.~\ref{geometry}~(b)).
The favorable energy of this deformation, 
$-f (R_\parallel-R)$,
is offset by surface energetics as well as a loss of
entropy, $S_{\rm d}$.  We estimate this entropy loss through the 
statistics of a Gaussian chain confined to a deformed volume.
In the long-chain limit, the free energy per monomer of
such a chain is isomorphic with the ground state energy
of a quantal particle confined to the same 
volume\cite{Lifshitz_RMP_1978}.
Treating the asymmetric boundary condition as a 
perturbation\cite{Landau_Lifshitz_QM}, we sum an infinite
class of terms in the ground-state expansion\cite{forthcoming},
obtaining
\begin{equation}
S_{\rm d} = S_{\rm sph}\left({2R_\parallel\over 3R}
+ {R^2\over 3 R_\parallel^2} - 1\right).
\label{entropy_deformation}
\end{equation}
Because $S_{\rm sph}\sim N^{2/3}$, 
the energy gained by a reasonable deformation
of the globule ($R_\parallel \sim N^{1/3}$), 
is comparable to $S_{\rm d}$
only for forces of magnitude $N^{1/3}$ or larger.
For long chains, the deformation achieved by stretching
forces $\sim T/a$ is therefore negligible.
We subsequently consider the globule to be undeformed,
and the globular free energy to be unaffected by stretching.

For sufficiently large stretching forces, the dominant state of
a homopolymer is an expanded coil (Fig.~\ref{geometry}~(d)).  
In contrast to the globule,
this state is characterized by extensive density fluctuations
and is strongly susceptible to deformation.
Considering short-ranged attractions between
monomers to be unimportant in this case, we model the extended coil
as a freely jointed chain.  The free energy of such a 
chain\cite{Grosberg_Khokhlov}, 
$F_{\rm c}(N) = -NT\ln{[\sinh{(fa/T)}/(fa/T)]}-TS_0$,
is sensitive to the magnitude of $f$.
At a given temperature $T<\theta$,
a phase transition occurs at a force sufficient
to lower the coil free energy below that of the globule.
(See inset of Fig.~\ref{phase_homo}.)
For $N^{2/3} \gg 1$, the phase boundary is given by 
\begin{equation}
B + \gamma N^{-1/3}
= -T\ln{[\sinh{(f a/T)}/(f a/T)]}.
\label{homo_boundary}
\end{equation}
The resulting phase diagram in the force-temperature
plane is shown in Fig.~\ref{phase_homo} for various $N$.
Here, we have taken $B=T-\theta$ (accurate near
$T=\theta$), which is 
unrealistic at low temperatures.
It is reasonable, however, that as temperature
decreases, stretching energy density 
of the strongly fluctuating coil
grows more quickly
than attractive energy density of the relatively placid globule.  
As a result, a reentrant coil phase appears at 
low temperature.
The ``coil'' in this case is a nearly straight chain
with small fluctuations in extension.
Computer simulations of strained
lattice heteropolymers\cite{Thir_PNAS_1999}
appear to support our prediction of reentrance.

Although the globule-coil transition is second-order
in the absence of stretching\cite{Grosberg_Khokhlov}, 
the force-induced transition
is here predicted to be first-order, since the average
chain extension is discontinuous at the phase boundary.
This result is certainly correct for $T<\theta$,
where globule and coil phases are distinct.
Near $T=\theta$ and $f=0$, however, our caricatures of these states
are oversimplified: Density fluctuations are not negligible in
the globule, and attractions are not negligible in the coil.
In this region of the phase diagram, 
extension increases smoothly with force, and the 
stretching transition is 
second-order.

In constructing a phase diagram for homopolymer stretching,
we did not consider 
chain structures that are
necklace-like (i.e., coexisting globule and coil regions, as in
Fig.~\ref{geometry}~(c)). 
Neglecting surface effects, the free energy per monomer
of a necklace lies between that of a globule and that of a coil.
The entropy of a necklace is augmented by the 
freedom of globular regions to reside 
anywhere along the chain
(provided they do not overlap).
For a single globular region, however, 
this additional ``translational'' entropy 
scales as $\ln{N}$, 
and is insufficient to overcome the 
$O(N)$ deficit in free energy to either globule or coil, even
for chains of modest length.  
In the case of many globular regions,
the gained entropy, $\sim N \ln{2}$, is considerable,
but does not compensate for the cost of presenting
an extensive surface to the solvent.

The stability and importance of necklace structures may be
qualitatively different for heteropolymers.  
In particular, fluctuations in local sequence composition
yield a preference for globule or coil that varies
along the chain.  We assess the strength of this
effect and its influence on stretching behavior, for a two-letter
random copolymer.  In this model, each monomer has two possible
identities, $\sigma_i = \pm 1$, perhaps denoting hydrophobic
and hydrophilic moieties.  For a given sequence $\{\sigma_i \}$,
the energy of a chain conformation defined by 
monomer positions
${\bf r}_i$ is
\begin{eqnarray}
{\cal H} &=& {\cal H}_0 + \Gamma \sum_{i \,\,{\rm exposed}} 
\sigma_i - {\bf f} \cdot ({\bf r}_N - {\bf r}_1),
\label{h_copolymer}
\\
{\cal H}_0 &=& \sum_{i,j=1}^N \delta({\bf r}_i - {\bf r}_j)
(B_0 + \chi \sigma_i \sigma_j), 
\end{eqnarray}
where $B_0$ is a mean attractive energy density stabilizing 
the globule at low temperatures.  We consider $\chi<0$,
so that attractions are strongest between monomers of
the same type. 
The second sum in Eq.~\ref{h_copolymer} includes only
monomers that are exposed to solvent.
For $\Gamma>0$, 
monomers of type $\sigma_i=-1$ are favorably solvated, 
while exposure of $\sigma_i=1$ monomers is unfavorable.
In Ref.~\cite{Shakh_PRE_1993}, 
the phase diagram was determined
for the model defined by ${\cal H}_0$
at the mean field level,
using the replica trick to average over 
random sequences of monomer 
types\cite{Shakh_biophch_1989,Binder_RMP_1986}.
The one-step replica symmetry breaking 
demonstrated in that work is consistent
with a suitably chosen random energy model.
In other words, the average thermodynamics are reproduced
by drawing $(a^3/v)^N$ energy levels at random from a 
distribution
$P(E) = (\pi N \Delta^2)^{-1/2} 
\exp{[-(E-\overline{E})^2/N\Delta^2]},$
with variance $\Delta=|\chi|\mu^2 \overline{\rho}$ and
mean $\overline{E}=B_0 N \overline{\rho}$.
Here, $v$ is the volume occupied by a monomer,
$\overline{\rho}\sim v^{-3}$ is the mean density in the globule core,
and $\mu$ is the variance of the monomer distribution.
Here we describe the effect of solvation
and stretching energetics on the effective distribution of 
energies\cite{forthcoming}.

For a compact, spherical globule with surface area $A$, 
the solvation contribution in Eq.~\ref{h_copolymer}
broadens the energy distribution, increasing 
$\Delta$ by $(\Gamma^2 \overline{\rho}/4 |\chi|)A/N$.
This result is obtained following the analysis of 
Ref.~\cite{Shakh_PRE_1993},
with the additional assumption
that the spatial pattern of
solvent-exposed monomers is independent of
compact chain conformation.
As in the case of the homopolymer,
we neglect the energetic contribution of globule deformation.
The resulting distribution of energies is dominated by
states in the interval
$\overline{E}-N^{1/2}\Delta < E < 
\overline{E}+N^{1/2}\Delta$.
At energies just below a critical value,
$E^*=\overline{E} - N\Delta(\ln{a^3/v})^{1/2}$,
the number of states is $O(1)$, while just above $E^*$
the number is  exponentially large.
The ground states of particular random sequences
are distributed narrowly about $E^*$\cite{Derrida_PRB_1981}.
Introducing solvation energetics thus lowers
the average ground state energy by an amount 
$(\Gamma^2 \overline{\rho}/ 4 |\chi|)
(\ln{a^3/v})^{1/2} A$.
If $\chi$ and $\Gamma$ 
are comparable in magnitude, this shift is a 
significant fraction of the energy gained by exposing 
only monomers with $\sigma_i=-1$ to the solvent.
The density of states around $E^*$ in this case
is sufficiently large that a ground state with favorable
solvation energetics may always be selected.

In an extended coil, nearly all monomers are exposed to solvent.
For sequences with fixed total composition, 
$\sum_{i=1}^N \sigma_i = 0$, thermodynamics of the coil state
are unaffected by solvation energetics.
We consider the heteropolymer coil to have essentially the same physics
as the homopolymer coil described above.

For necklace structures of a random heteropolymer,
free energy depends on the positions of globular regions
along the chain.  In effect, these globules move in 
a random potential generated by sequence disorder.
(See Fig.~\ref{random_potential}.)
The scale and correlation length of this
potential are determined by the size of the globule, and by 
statistics of the sequence.
Although the total composition of the chain is fixed,
a globular region with $n<N$ monomers, situated at a given
point on the chain, has in general an excess of one monomer
type:
\begin{equation}
q\equiv {1\over n}\sum_{i \in {\rm globule}} \sigma_i \neq 0.
\end{equation}
Because the local composition
$q$ is a sum of many independent random variables, its
distribution is Gaussian, with variance $\mu n^{-1/2}$.
This distribution of sequence compositions leads 
to localization of globules along the chain.

At a particular location, a globular region has an apparent
distribution of monomer types
that is modified according to the local value of $q$,
$p(\sigma_i;q)\propto \exp{[-(\sigma_i-q)^2/2\mu^2 (1-q^2)]}$.
The ground state energy of the globule at this location,
determined from the random energy model described above, is
\begin{equation}
E^* = (B_0 - |\chi|q^2) \overline{\rho} n
 - \Gamma q \overline{\rho} n^{2/3}
 - \left(\ln{a^3 \over v}\right)^{1/2}
\left[2|\chi|\mu^2 (1-q^2)\overline{\rho} n
 + {\Gamma^2 \overline{\rho} \over 4|\chi|}n^{2/3}
 \right].
\end{equation}
Since local composition is not fixed for a heteropolymer,
$E^*$ is a random variable for different globule locations.
But its fluctuations are weak:
$\langle (\delta E^*)^2  \rangle^{1/2}
= \Gamma \overline{\rho}\mu n^{1/6} + O(n^0)$,
where angled brackets denote an average over the
distribution of $q$.

For a single globular region incorporating $n$ monomers,
the remaining $N-n$ monomers of the necklace belong
to coil regions.  When the globule resides at 
a given chain location 
with composition $q$, the complementary composition in 
coil regions is $-q$.
Consequently, the solvation energy 
of these fully exposed regions,
$-\Gamma n q$, is also a random variable,
with fluctuations of magnitude $\Gamma \mu n^{1/2}$.  
These relatively strong fluctuations
in coil solvation energy
establish the scale of the random potential
for globule motion along the chain.
Below a critical temperature $T_{\rm c}$, the globule will
become localized in the deepest minimum of this
potential.  Following Derrida's analysis of randomly 
distributed energies\cite{Derrida_PRL_1980}, 
we find that $T_{\rm c} = \Gamma \mu n^{1/2}/
\sqrt{2 s}$, 
and that the free energy due to globule motion is
\begin{equation}
F_{\rm rand} = 
\cases{
-sT[1+({T_{\rm c} \over T})^2], & $T>T_{\rm c}$\cr
-2sT_{\rm c}, & $T\leq T_{\rm c}.$\cr
}
\label{hetero_necklace}
\end{equation}
In Eq.~\ref{hetero_necklace}, $s$
is the logarithm of the
number of possible globule locations.  
The number of statistically independent
locations is approximately a factor of $n$ smaller, 
so that $s$ is appropriately
renormalized, $s \simeq \ln{(N/n)-1}$.
For $T \gg T_{\rm c}$, randomness of the 
potential is irrelevant, and the homopolymer result,
$\sim T\ln{N}$ is recovered.

The analysis for a single globule is readily generalized 
for a necklace with several globular regions.  In this case,
$e^s \simeq n^{-M}
(N-n+1)(N-2n+1)\ldots(N-Mn+1)/M!$, where $M$ is
the number of globules.
For small globules, $n=O(1)$, the minimum value of 
$F_{\rm rand}$ is obtained for $M=O(N)$, giving
$\min{F_{\rm rand}} = O(N^{1/2})$.
For large globules, $n=O(N)$, $M$ is necessarily $O(1)$,
and again $\min{F_{\rm rand}} = O(N^{1/2})$.
Globule sizes are thus expected
to be distributed broadly, with a
preference for a small
number of large globules dictated by surface tension.
Due to the $O(N^{1/2})$ stabilization provided
by globule localization, a phase rich in necklace
structures covers an appreciable
range of force and temperature for $N=10^2$ 
(Fig.~\ref{phase_hetero}~(a)).  
Indeed, simulated stretching of a short, nearly random
heteropolymer involves
partially extended structures\cite{Thir_PNAS_1999}.
For $N=10^3$ (Fig.~\ref{phase_hetero}~(b)), the necklace
phase covers a much smaller region of the phase diagram,
due to the extensive free energy 
cost of mixing
globule and coil structures.
Although necklaces become unstable as $N\rightarrow\infty$,
our results suggest that they may be prevalent
for macromolecules of finite size.

From our results for uncorrelated sequence statistics,
we may deduce the basic effects of introducing correlations.
For ``blocky'' sequences, in which monomers
within a correlation length $\xi$ are likely to be
of the same type, fluctuations in local composition
are large.  Specifically, 
$\langle (\delta q)^2 \rangle = O(n^0)$ for $n<\xi$.
If $\xi$ scales with chain length
(or if sequence correlations decay algebraically with 
distance along the chain), 
the resulting free energy
contribution due to localization is $O(N)$, and necklace
structures persist as $N\rightarrow\infty$.
By contrast, sequences that are anticorrelated on a scale
$\xi$ exhibit small fluctuations in local composition.
If $\langle (\delta q)^2 \rangle = O(n^{-1/2})$,
fluctuations in solvation energy are $O(1)$, and the stretching
behavior will be homopolymeric.

The equilibrium behavior of strained polymers described in this
Letter is not directly related to the intrinsically dynamical,
nonequilibrium response measured in experiments.
Nonetheless, it is reasonable to associate the equilibrium
stability of various structures with their kinetic accessibility.
In particular, when necklaces are stable over a wide
range of stretching forces, the kinetic transition 
from compact globule to extended coil is likely not sharp.  
Instead, we expect 
that the chain visits an ensemble of necklace structures as it
passes through 
the transition region.  The breadth we have predicted for this
ensemble in the case of random heteropolymers suggests
that the restoring force to applied strain should exhibit 
large fluctuations as the chain is stretched.  Such a scenario
has in fact been observed for the protein 
barnase\cite{Clarke_communication}.
The observation that certain protein domains
do undergo sharp stretching transitions is thus an indication 
of evolutionary design for mechanical strength,
as is reflected by the roles of such proteins in cell adhesion
and muscle elasticity.

\newpage
\bibliographystyle{prsty}

\begin{thebibliography}{10}

\bibitem{Fernandez_nat_1998}
A.~F. Oberhauser, P.~E. Marszalek, H.~P. Erickson, and J.~M. Fernandez, Nature
  {\bf 393},  181  (1998).

\bibitem{Gaub_1998}
M. Rief, M. Gautel, A. Schemmel, and H. Gaub, Biophys. J. {\bf 75},  3008
  (1998).

\bibitem{Fernandez_nat_1999}
P.~E. Marszalek {\it et~al.}, Nature {\bf 402},  100  (1999).

\bibitem{Fernandez_PNAS_2000}
H. Li {\it et~al.}, Proc. Nat. Acad. Sci. {\bf 97},  6527  (2000).

\bibitem{Bustamante_PNAS_2000}
G. Yang {\it et~al.}, Proc. Nat. Acad. Sci. {\bf 97},  139  (2000).

\bibitem{Wolynes_PNAS_1999}
N.~D. Socci, J.~N. Onuchic, and P.~G. Wolynes, Proc. Nat. Acad. Sci. {\bf 96},
  2031  (1999).

\bibitem{Schulten_proteins_1999}
H. Lu and K. Schulten, Proteins: Struc. Func. Genet. {\bf 35},  453  (1999).

\bibitem{Grosberg_Khokhlov}
A.~Y. Grosberg and A.~R. Khokhlov, {\em Statistical Mechanics of Chain
  Molecules} (American Institute of Physics, Woodbury, NY, 1994).

\bibitem{Lifshitz_RMP_1978}
E.~M. Lifshitz, A.~Y. Grosberg, and A.~R. Khokhlov, Rev. Mod. Phys. {\bf 50},
  683  (1978).

\bibitem{Landau_Lifshitz_QM}
L.~D. Landau and E.~M. Lifshitz, {\em Quantum Mechanics: Non-relativistc
  Theory} (Butterworth-Heinemann, Boston, 1991).

\bibitem{forthcoming}
P.~L. Geissler and E.~I. Shakhnovich, in preparation.

\bibitem{Thir_PNAS_1999}
D.~K. Klimov and D. Thirumalai, Proc. Nat. Acad. Sci. {\bf 96},  6166  (1999).

\bibitem{Shakh_PRE_1993}
C.~D. Sfatos, A.~M. Gutin, and E.~I. Shakhnovich, Phys. Rev. E {\bf 48},  465
  (1993).

\bibitem{Shakh_biophch_1989}
E.~I. Shakhnovich and A.~M. Gutin, Biophys. Chem. {\bf 34},  187  (1989).

\bibitem{Binder_RMP_1986}
K. Binder and A.~P. Young, Rev. Mod. Phys. {\bf 58},  801  (1986).

\bibitem{Derrida_PRB_1981}
B. Derrida, Phys. Rev. B {\bf 24},  2613  (1981).

\bibitem{Derrida_PRL_1980}
B. Derrida, Phys. Rev. Lett. {\bf 45},  79  (1980).

\bibitem{Clarke_communication}
J. Clarke, 2001, private communication.

\end{thebibliography}

\newpage
\begin{figure}
\caption{
Possible states of a strained polymer:
(a) compact, spherical globule;
(b) compact globule, deformed from a spherical geometry, with
extension $2 R_\parallel$ in the direction of stretching;
(c) necklace of alternating compact and non-compact regions;
and (d) fully non-compact, extensible coil.
}
\label{geometry}
\end{figure}

\newpage
\begin{figure}
\caption{
Phase diagram of a homopolymer subject to an applied force,
$f$, on the end-to-end distance, as estimated by 
Eq.~\ref{homo_boundary}.  The boundary between globule
and coil phases is drawn for $N=10^2$ (dot-dashed line),
$N=10^3$ (dotted line), $N=10^4$ (dashed line), and
$N\rightarrow\infty$ (solid line).  
These results are obtained for 
a surface energy density that is comparable
to monomer interactions,
$\gamma/\theta \approx 1$.
Inset: Schematic picture of the influence
of stretching on the free energy of extension for
$T<\theta$.
As force is increased ($f_1<f_2<f_3<f_4$),
the coil state first becomes metastable
(represented by the local free energy minimum 
at large $R$) and then stable.
Free energy of the globular state
(represented by the local minimum at small $R$)
is not sensitive to $f$.
}
\label{phase_homo}
\end{figure}

\newpage
\begin{figure}
\caption{
Motion of a globular region along a random heteropolymer.
A sequence of monomer identities is represented
schematically by black and white circles on the chain.
Different sequence locations, $x$, of the globule
result in different compositions (here, fractions of
black and white circles) of the globule and of the
extended coil regions.
Consequently, the ground state energy of the globule
and solvation energy of the coil depend on $x$.
Because the monomer identities are independent
random variables,
the globule effectively experiences a random,
one-dimensional potential $u$ for translation along 
the sequence.
}
\label{random_potential}
\end{figure}

\newpage
\begin{figure}
\vspace*{-2cm}
\caption{
Phase diagram of the random copolymer defined by 
Eq.~\ref{h_copolymer} for (a) $N=10^2$ and (b) $N=10^3$,
estimated using mean-field arguments described in the text.
Shading denotes regions dominated by necklace-like structures.
Boundaries of these regions are taken to be
chain compositions of 75\% globule (lower dashed lines) and
25\% globule (upper dashed lines).
The solid line denotes a chain composition of 50\% globule,
i.e., $N/2$ monomers belong to globular regions, and $N/2$
belong to coil regions.
The chain is assumed to be flexible on the scale
of monomer size, 
$\overline{\rho} a^3 \approx 1$.
}
\label{phase_hetero}
\end{figure}

\end{document}